\documentclass[preprint,showpacs,prl,12pt,onecolumn]{revtex4}
\usepackage{amssymb}

\input{tcilatex}

\begin{document}

\preprint{HEP/123-qed}
\title[Self-consistent Green function approach ]{Self-consistent Green
function approach for calculations of electronic structure in transition
metals}
\author{N.E.Zein and V.P.Antropov}
\affiliation{Ames Lab, Ames, IA, 50011}
\keywords{Many body Green function, electronic structure}
\pacs{71.28.+d, 71.25.Pi, 75.30.Mb}

\begin{abstract}
We present an approach for self-consistent calculations of the many-body
Green function in transition metals. The distinguishing feature of our
approach is the use of the one-site approximation and the self-consistent
quasiparticle wave function basis set, obtained from the solution of the
Schrodinger equation with a nonlocal potential. We analyze several sets of
skeleton diagrams as generating functionals for the Green function
self-energy, including GW and fluctuating exchange sets. Their relative
contribution to the electronic structure in 3d-metals was identified.
Calculations for Fe and Ni revealed stronger energy dependence of the
effective interaction and self-energy of the d-electrons near the Fermi
level \ compared to s and p electron states. Reasonable agreement with
experimental results is obtained.
\end{abstract}

\maketitle

Density-functional theory (DFT), in particular the local-density
approximation (LDA), has proven to be a rather successful \textit{ab-initio}
approach to describe physical properties of many materials. Nevertheless
numerous applications of this method have revealed a number of shortcomings
related to the inadequate treatment of both excited (energy gap) and
strongly correlated states. For a long time the many-body Green function
(GF) approach was considered as a possible \textit{ab-initio} alternative to
DFT, but its current applicability is usually restricted to the homogeneous
electron gas (HEG) model\cite{holm1,holm3,godby} or semiconductors in the GW
approximation. However, realization of the full self-consistent GW scheme is
complicated due to enormous computational difficulties. As a result, for the
transition metals (TM) the GW approach was applied only to Ni\cite%
{arya1,arya2} and these calculations used the wave functions obtained within
the LDA as the basis set, i.e. they were not self-consistent. Thus, it is
quite unlikely that a universal GF method can be formulated and tailoring
the approximations to specific properties of a given class of materials is
desirable. The localized character of the d-wave function in TM often
implies the well-known one-site approximation\cite{kot} which in real
systems acquires certain material specific features.

Already the calculations of Refs.\cite{arya1,arya2} revealed a strong
energy-dependence of the effective interaction in TM, even in the vicinity
of the Fermi level. For example in Ni the effective Coulomb interaction $W$
is nearly zero at 6 eV and increases to an unscreened value of 20-25 eV at
20 eV. Another important observation is related to a proper intra-atomic
scale treatment. Elementary analysis shows that the $d$- function in bcc Fe
is significantly altered at a distance as small as 0.8 a.u., while $s$ and $p
$ wave functions are altered at distances of about the Wigner-Seitz radius ($%
\sim $ 2.6 a.u.). The \textquotedblleft local\textquotedblright\ screening
length, $\lambda =\varkappa ^{-1}$ $(\varkappa ^{2}(r)=4\pi e^{2}\Pi
_{0}(\rho (r)),$ with $\Pi _{0}(\rho (r))=(3/\pi \rho )^{1/3}/\pi ),$ which
varies from 0.5 a.u. in the region of the maximum of the $d-$wave function
to a value of 1.0 a.u. at the atomic sphere (AS) boundary, provides an
initial justification for the one-site approximation and suggests that the
correct material-specific implementation must include a spatially resolved
representation of the polarization operator (PO) and other two-channel
operators, at least inside the AS. We believe that the energy dependence and
intra-atomic resolution are both crucial for the quantitative description of
TM and must be included explicitly in any reliable technique. On the other
hand, we will show below that this approximation can describe the essential
physics of TM, greatly reducing the computational efforts.

In the present paper we incorporate the effects described above into the
self-consistent GF technique using the quasiparticle wave function basis set
and the Luttinger-Ward functional\cite{LW} (LWF) approach for the
self-energy calculations. We tested several sets of diagrams, mainly from
the fluctuating-exchange subset (FLEX)\cite{flex1}. All diagrams beyond
these sets were taken in the local approximation. The use of the LWF
guarantees the equivalence of one-particle properties calculated with the GF
and with the corresponding total energy variation\cite{baym}. The LWF
formalism naturally leads to the integration over the imaginary axis,
greatly improving the numerical accuracy of the integration. On the
imaginary axis both the GF and PO quickly reach their quasiclassical limits,
are sufficiently smooth, and are determined by the local potential only\cite%
{KS}. The rotation of the integration contour leads to a very convenient
separation of the structural and local density dependent degrees of freedom
for both self-energy and total energy\cite{khodel1,zein}. Also, in contrast
with the numerical method\cite{godby} we analytically select the
contributions from both the GF cut on the real axis and the integration
along the imaginary axis. An energy linearization of the quasiparticle wave
function similar to the one in linear density functional methods was used.

In terms of an exact GF the thermodynamic potential is written as\cite{LW} 
\begin{equation}
\Omega =-Tr\left\{ \ln \left[ \Sigma -G_{0}^{-1}\right] \right\} -Tr\Sigma
G-\Phi   \label{LWF}
\end{equation}%
where $\Sigma $ is the self-energy, G and G$_{0}$ are exact and 'bare' GF
and $\Phi $ is the Luttinger generating functional, which is represented by
the set of skeleton graphs. Minimizing $\Omega $ over $G$ one can obtain\cite%
{LW} 
\begin{equation}
\Sigma =\frac{\delta \Phi }{\delta G}
\end{equation}%
The expressions for $\Phi $ for the most important two-particle (hole)
channels are summarized in Ref.\onlinecite{flex1}. The set from Fig.1b with
'bare' loops corresponds to the GW-approximation. After variation of $\Phi $
over the GF, which we consider as a variational variable, one can obtain the
usual GW expression for the self-energy: 
\begin{equation}
\Sigma (\varepsilon )=-\int G(\varepsilon -\omega )V_{c}\Pi (\omega
)W(\omega )\frac{d\omega }{2\pi i}  \label{SGM}
\end{equation}%
where $V_{c}$ is the Coulomb interaction and $W$ is the effective
interaction $W=V_{c}/(1+V_{c}\Pi )$. It is convenient to rotate the contour
of integration in Eq.(\ref{SGM}) in the complex plane, using the fact that $%
\Pi (z)$ has no singularities in the first and third quadrants\cite%
{galitsky,mahan}, whereas $G$ has a singularity (cut) in the third quadrant
for $\varepsilon <$ $E_{F}$ and in the first one for $\varepsilon >$ $E_{F}$%
. In the case of the quasiparticle the singularities of the GF are simple
poles. After rotation the only additional contribution comes from the cut of
the GF (Fig 1e), and for $\varepsilon <$ $E_{F}$ this contribution is 
\[
\Sigma _{ca}^{p}=\int_{\varepsilon }^{0}g(\omega ^{\prime })\Pi (\varepsilon
-\omega ^{\prime })V_{c}W(\varepsilon -\omega ^{\prime })d\omega ^{\prime
}\quad \varepsilon <0,
\]%
where $g(\omega )=-\mathop{\rm Im}G(\omega )\mathrm{sgn}\omega /\pi $. For
the term corresponding to the integration along the imaginary axis we obtain 
\[
\Sigma _{ca}^{i}=\int\limits_{-\infty }^{\infty }d\omega ^{\prime
}\int\limits_{0}^{\infty }\frac{d\omega }{2\pi i}g(\omega ^{\prime })\frac{%
2(\varepsilon -\omega ^{\prime })}{(\varepsilon -\omega ^{\prime
})^{2}+\omega ^{2}}\Pi (i\omega )V_{c}W(i\omega )
\]%
At $E_{F},$ $\varepsilon =0$ and the pole term disappears. The term $\Sigma
_{ca}^{i}$ is usually negative in the HEG model. On the real axis $\Pi
(\omega )$ has an imaginary part, which is important for accurate
calculations of $W(\omega )$ in TM. The summation over \textbf{k} (one-site
approximation) transforms all the above formulas into matrix equations on
the local wave function basis, greatly reducing the computational efforts.
The full on-site GF $G\left( \mathbf{r},\mathbf{r}^{\prime },\varepsilon
\right) =\sum_{\mathbf{k}}\left( G_{0}^{-1}\left( \mathbf{r},\mathbf{r}%
^{\prime },\varepsilon ,\mathbf{k}\right) -\Sigma \left( \mathbf{r},\mathbf{r%
}^{\prime },\varepsilon \right) \right) ^{-1}$ was obtained
self-consistently from Eq.(\ref{SGM}). Due to the large value of $\partial
\Sigma /\partial \varepsilon $ in TM, the final $G$ differs significantly
from the initial $G_{0}$. The quantity $\Pi \left( \mathbf{r},\mathbf{r}%
^{\prime },\varepsilon \right) =\int G\left( \varepsilon \right) G\left(
\varepsilon +\omega \right) d\varepsilon /2\pi i$ was also obtained
self-consistently with the full $G$.

Let us discuss the choice of skeleton graphs for the LWF. According to
perturbation theory with Coulomb parameter $\alpha =e^{2}m/p_{f}$ the
exchange diagram (Fig 1a) gives the largest contribution. The set of empty
bubble diagrams (Fig.1b) gives the next term $\alpha ^{2}\ln 1/\alpha $. The
summation of these terms is necessary to take into account the long-distance
character of the Coulomb interaction. In addition, the proper treatment of
the ''dressed''\ GF in\ the bubble approximation must include vertex
corrections\cite{holas} (Fig.1b). The next term (Fig.1c) is just the
exchange diagram in the second-order approximation, which has an order of $%
\alpha ^{2}$. It can be important in magnets, as the corresponding
self-energy depends only on the GF with the same spin, while in bubble
diagrams both spins are averaged. In principle, the Fig.1c diagram is
already included in the set, if the bubble diagrams with vertex corrections
are considered. But it seems desirable to include also the ladder sequence
of Fig.1d, as it is also highly spin dependent. This set does not contain
the first ($\alpha ^{2}$) term\cite{flex1}. Such a sum is reduced by the
effective screening of the Coulomb potential, which in turn strongly depends
on the energy. The static value of the effective interaction is rather
small, but it quickly increases as a function of energy (Fig.2b). We
evaluated this sum of lattice-type T-matrix diagrams (Fig.1d) with the
effective interaction replacing the energy dependent potential by the
averaged static interaction $\zeta V_{c}$. The parameter $\zeta $ was chosen
in such a way that the value of the diagrams on Fig.1c obtained with both
energy dependent and $\zeta V_{c}$ type of interactions were the same. The
value of $\zeta $ turned out to be $\sim $ 0.35.

The diagrams discussed above are also the leading diagrams from the point of
view of the one-site approximation. The number of d-electrons with various
angular moment projections $m_{z}$ is approximately the same in both Ni and
Fe. If we take into account that the main contribution in the Coulomb
interaction, expanded in spherical harmonics $V(\mathbf{r-r^{\prime }}%
)=\sum_{L}Y_{L}(\widehat{r})Y_{L}(\widehat{r}^{\prime })(r_{<}/r_{>})^{l}$
is from the term with $l=0$, then, with this accuracy, $m_{z}$ is conserved
and one can classify various diagrams with the parameter $1/N=1/(2l+1)$. For
d electrons this is a small parameter and the main terms with $1/N$ are
again FLEX diagrams\cite{foerster}. In summary, we believe, that the set of
diagrams in Fig.1 is the minimal set which must be included in the
calculations. On the other hand, our calculations revealed that it can be
sufficient for description of magnetic properties and electronic structure
of 3d metals.

The rotation of the integration contour allows us to deal only with the
states lying in the vicinity of $E_{F}$\cite{zein}, as the PO is smooth and
decreases when Im $\varepsilon $ is increased, contrary to its behavior on
the real axis. This is shown in Fig.2a,b for ferromagnetic Fe. For the
states near $E_{F}$ we can use the GF constructed with low lying LMTO states 
$\Psi _{\mathbf{k}}^{\nu }(\mathbf{r})=\sum_{L}a_{\mathbf{k}}^{\nu \alpha
}\phi _{l}^{\alpha }(r)Y_{L}(\widehat{n})$, where $\phi _{l}^{0}(r)$ and $%
\phi _{l}^{1}(r)$ are the corresponding solutions of equations%
\begin{equation}
\widehat{G}_{0l}^{-1}\phi _{l}^{0}(r)=(\varepsilon -\widehat{T})\phi
_{l}^{0}(r)-\int V(\mathbf{r,r^{\prime }})\phi _{l}^{0}(r^{\prime
})d^{3}r^{\prime }=0  \label{sch}
\end{equation}%
\begin{equation}
\widehat{G}_{0l}^{-1}\phi _{l}^{1}(r)=-\phi _{l}^{0}(r)-\int \frac{\partial
\Sigma (\varepsilon _{l},\mathbf{r,r^{\prime }})}{\partial \varepsilon _{l}}%
\phi _{l}^{0}(r^{\prime })d^{3}r^{\prime },  \label{dsch}
\end{equation}%
with $V(\mathbf{r,r^{\prime }})=\Sigma (\varepsilon _{l},\mathbf{r,r^{\prime
}}).$ $\varepsilon _{l}$ is the center of gravity of the band with orbital
moment $l$ and $a_{\mathbf{k}}^{\nu 0}$ are the eigenvectors of the
generalized eigenvalue problem with the Hamiltonian 
\begin{equation}
H=H_{0}+\Sigma _{l}(\varepsilon )-\Sigma _{l}(\varepsilon _{l})-(\varepsilon
-\varepsilon _{l})\dot{\Sigma}_{l}(\varepsilon _{l})
\end{equation}%
which is a matrix in the space of LMTO states $\phi +h\dot{\phi}$, where $%
h_{LL^{\prime }}$ are the usual LMTO coefficients. This procedure takes into
account the energy-independent exchange potential $\Sigma _{x}=1/2\pi i\int
V_{c}\mathrm{Im}Gd\varepsilon $ exactly.

The important problem in the nonlocal calculations is the inclusion of the
valence-core interaction. In our method we included the core-valence
exchange term exactly, whereas the core-valence correlation term was added
using the approximation developed in Ref.\cite{martin}. The latter term in
TM such as Fe, Ni or Cu is small. The set of Eqs.(\ref{sch}),(\ref{dsch})
was solved iteratively\cite{zein}. The integral equation for the screened
interaction $D({\omega ,\mathbf{r,r^{\prime }}})$ was solved using the
product basis introduced in Ref.\cite{arya}.

The procedure above was applied to the 3d transition metals Fe and Ni. The
main contribution to the self-energy comes from the exchange diagram (Fig
1a), but the pure exchange approximation produces too large a
self-consistent magnetic moment $M$ ($M\sim $ 3.05$\mu _{B}$ in Fe (see also
Ref.\cite{kotani} )). The bubble diagrams screen the Coulomb interaction,
either too strongly ($M\sim $ 1.95$\mu _{B}$ in Fe) or too weakly ($M\sim $
0.73$\mu _{B}$ in Ni). Vertex corrections only slightly modify this result,
uniformly reducing the interaction for both spins. The conclusions in
general agree with GW results for the HEG\cite{holm1,holm3,godby}, though
the importance of the exchange T-matrix diagram (Fig 1d) is rather intrinsic
to TM\cite{flex1,Tmatrix}. With all considered diagrams the calculated
equilibrium magnetic moments (2.04$\mu _{B}$ in Fe and 0.65$\mu _{B}$ in Ni
) are close to LDA values (2.15$\mu _{B}$, 0.62$\mu _{B}$) and experimental
values (2.08$\mu _{B}$, 0.59$\mu _{B}$).

The GF calculations naturally provide valuable information about the
renormalization and damping of the electron spectra which are absent in LDA.
In Fig.3 we present the energy dependence of the real ($\Sigma ^{R}$) and
imaginary ($\Sigma ^{I}$) parts of the self-energy $\Sigma _{2a}$. $\Sigma
^{R}\left( \omega \right) $ is linear in a wide range of energies,
supporting the concept of well-defined quasiparticles. On the other hand the
renormalization factor $Z_{l}=1/(1-\partial \Sigma /\partial \varepsilon )$
strongly depends on the orbital number $l$. For $s$ and $p$ electrons $Z_{l}$
is about 0.96 and is in good agreement with the HEG estimations ($Z\sim
1-0.04e^{2}m/p_{f}$ \cite{galitsky}). But for $d$ electrons this factor is
about 0.6 in Ni and 0.7 in Fe, pointing out severe many-body effects. The
value $Z_{d}$ was calculated self-consistently and in Ni it approximately
coincides with the value calculated with a non-self-consistent (using LDA
wave functions) method\cite{arya1}.

Previous studies revealed that the most significant differences between the
LDA and experimental results are the satellite peak below the quasiparticle
band (approximately 6-8 eV below $E_{F}$) and\ a larger value of the density
of states (DOS) at the Fermi level $N(E_{F})$. As a result, the overall
experimental bandwidth (7-8 eV in Fe and 6-7 eV in Ni) is larger than what
is found in the LDA along with the larger values of $N(E_{F})$. Extracted
from the linear term in the heat capacity, the experimental DOS values are
57 st/a.u. in Fe and 81 st/a.u. in Ni. Though a considerable amount of this
value can be ascribed to the electron-phonon interaction, they are still
much larger than the `bare' LDA values in Fe (30 st/a.u.) and in Ni (48
st/a.u.).

The small value of $Z_{d}$ obtained above leads to a narrowing of the
quasiparticle band. The rest of the one-electron states take part in the
formation of the satellite structure, which usually appears when the
hole-hole interaction is taken into account\cite{Tmatrix}. Consequently our
DOS (Fig.4) is also wider compared to the LDA DOS and has a bump below the
quasiparticle DOS. The value of the DOS at the Fermi level is even lower
than in the LDA, though the value of coefficient $\gamma $, which is $1/Z$
times larger than the DOS, is approximately the same as in LDA. Both values
are still lower than the experimental values. It can be due to the high
sensitivity of the Stoner splitting in ferromagnets (and in turn the value
of $N(E_{F})$) to all the approximations used or possible narrowing of the
hopping between neighboring sites in the spirit of the Gutzwiller
approximation or the Dynamical Mean Field Theory, which is omitted in our
approximation.

In conclusion, we have proposed a new self-consistent version of the GF
approach, which uses the quasiparticle wave functions as a basis set. We
found that self-consistent GW approach produces the leading contribution to
the electronic structure and magnetic properties of TM, whereas the addition
of fluctuating exchange diagrams overall slightly corrects this result
improving comparison with experiment. While the self-consistent
renormalization factors $Z_{l}$ for $s$ and $p$ electrons in Ni and Fe are
close to the estimations obtained from the HEG, $Z_{d}$ is much smaller
(0.6-0.7). The values of bandwidth and density of states at the Fermi level
are in reasonable agreement with experiment. The proposed technique can be
naturally used for total energy calculations. In summary we believe that the
proposed technique can be considered as a practical \textit{ab-initio}
alternative to modern DFT methods with much wider range of applicability.

Authors would like to acknowledge G.Kotliar, D.Lynch, S.Savrasov and M. van
Schilfgaarde for useful discussions and comments. This work was carried out
at the Ames Laboratory, which is operated for the U.S. Department of Energy
by Iowa State University under Contract No. W-7405-82. This work was
supported by the Division of the Office of Basic Energy Sciences of the U.S.
Department of Energy and by the Russian Fund for Basic Research under Grant
00-15-96709.

\section{Figure captions}

Fig 1 The diagram representation of: a) exchange energy, (b) ''bubble''\ set
with vertex corrections for the polarization operator, (c) exchange diagram
of second order (d) exchange T-matrix set. The integration contour and pole
positions in the complex energy plane are shown in (e).

Fig 2. The energy dependence of the real and imaginary parts of the
polarization operator with $l=2$ on the real axis (a). The same for the
polarization operator and effective interaction $V_{eff}=%
\sum_{l}D_{l}Z_{l}/Z $ on the imaginary axis (b).

Fig 3. The energy dependence of the partial components of self-energy in
ferromagnetic Fe: \ (a) the real part and \ (b) absolute value of imaginary
part.

Fig 4. The density of states in bcc Fe for the majority (solid) and minority
(dashed) spin states.


\bigskip

\begin{thebibliography}{99}
\bibitem{holm1} B.Holm and U. von Barth, Phys.Rev. \textbf{B57},2108 (1998)

\bibitem{holm3} B.Holm and F.Aryasetiawan, Phys.Rev. \textbf{B62},4858 (2000)

\bibitem{godby} P.Garcia-Gonzalez and R.W.Godby, Phys.Rev. \textbf{B63}%
,075112 (2001)

\bibitem{arya1} F.Aryasetiawan. Phys.Rev. \textbf{B46},13051 (1992)

\bibitem{arya2} M.Springer and F.Aryasetiawan, Phys.Rev. \textbf{B57}, 4364
(1998)

\bibitem{kot} A.Georges, G.Kotliar, W.Krauth, and M.J.Rozenberg,
Rev.Mod.Phys. \textbf{68}, 13 (1996)

\bibitem{LW} J.M.Luttinger and J.C.Ward, Phys.Rev. \textbf{118}, 1417 (1960)

\bibitem{flex1} N.E.Bickers and D.J.Scalapino, Ann. of Phys. \textbf{193},
206, (1989)

\bibitem{baym} G.Baym and L.Kadanoff, Phys.Rev. \textbf{124}, 287, (1961),
G. Baym, Phys.Rev. \textbf{127}, 1391, (1962)

\bibitem{KS} W.Kohn and L.J.Sham, Phys.Rev. \textbf{137A}, 1697 (1965)

\bibitem{khodel1} V.A.Khodel and E.E.Saperstein, Phys.Lett. \textbf{B36},
429 (1971)

\bibitem{zein} N.E.Zein and V.P.Antropov, J.Appl.Phys. \textbf{83}, 7314
(2001)

\bibitem{galitsky} V.M.Galitsky.\textit{\textquotedblright Collected papers
in theoretical physics\textquotedblright , }Moscow, Nauka, 1983, p.134

\bibitem{mahan} G.D.Mahan.\textit{\textquotedblright Many-particle
physics\textquotedblright ,} Plenum Press, 1990

\bibitem{holas} A.Holas, P.K.Aravind, K.S.Singwi, Phys.Rev. \textbf{B20},
4912 (1979)

\bibitem{foerster} D.Foerster, Phys.Rev. \textbf{B61},5066 (2000)

\bibitem{martin} E.L.Shirley and R.M.Martin, Phys.Rev. \textbf{B47}, 15413
(1993)

\bibitem{arya} F.Aryasetiawan and O.Gunnarson, Phys.Rev. \textbf{B49,} 16214
(1994)

\bibitem{kotani} T.Kotani, J.Phys.Cond.Mat. \textbf{10}, 9241 (1998); ibid.%
\textbf{12}, 2413 (2000)

\bibitem{Tmatrix} F.Manghi, V.Bellini, J.Osterwalder, T.J.Kreutz, P.Aebi,
C.Arcangeli, Phys.Rev. \textbf{B59}, R10409 (1999)
\end{thebibliography}
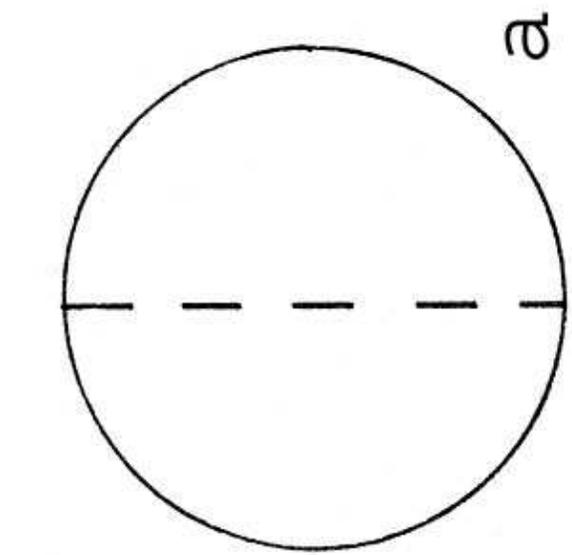
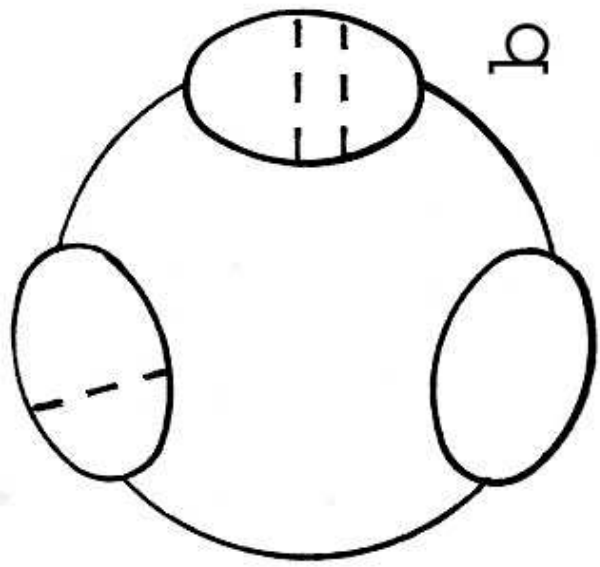
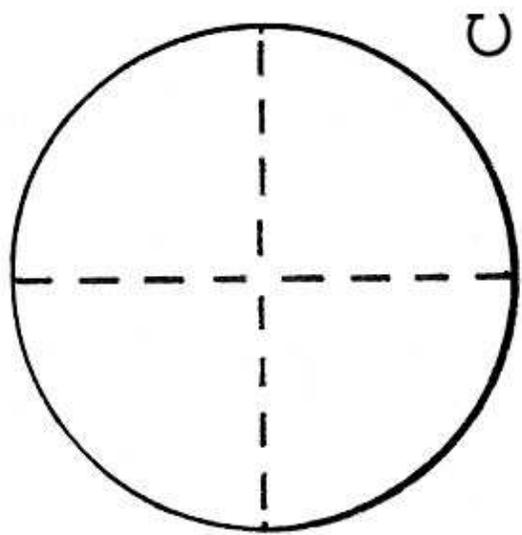
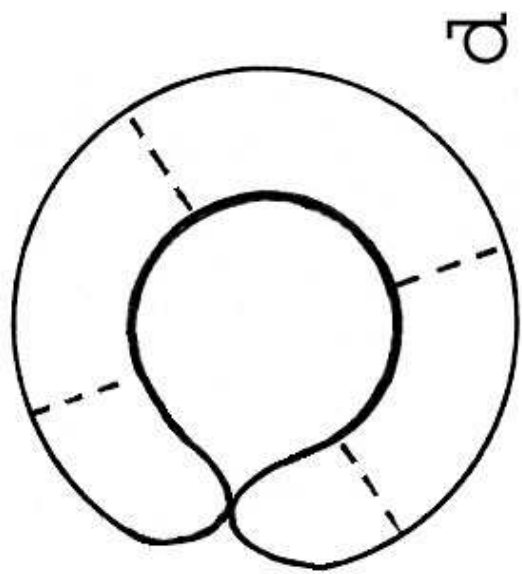
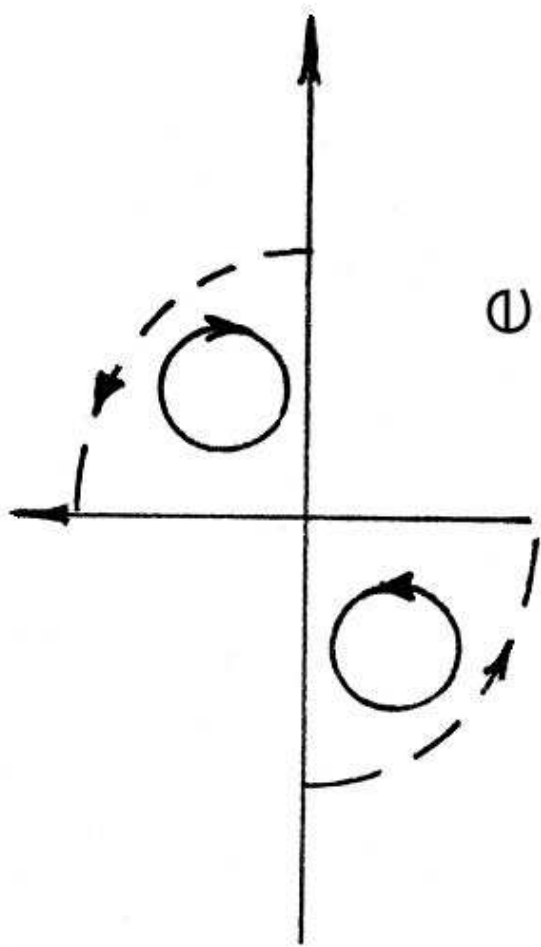
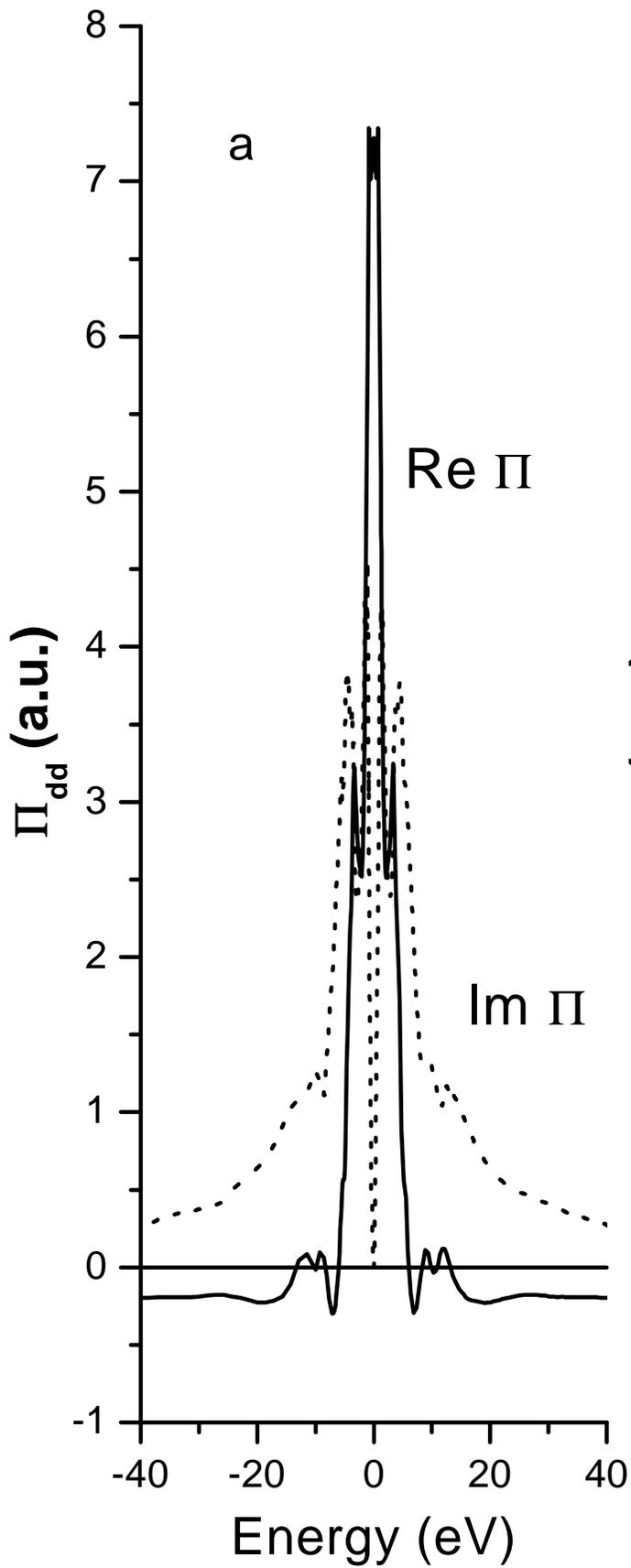
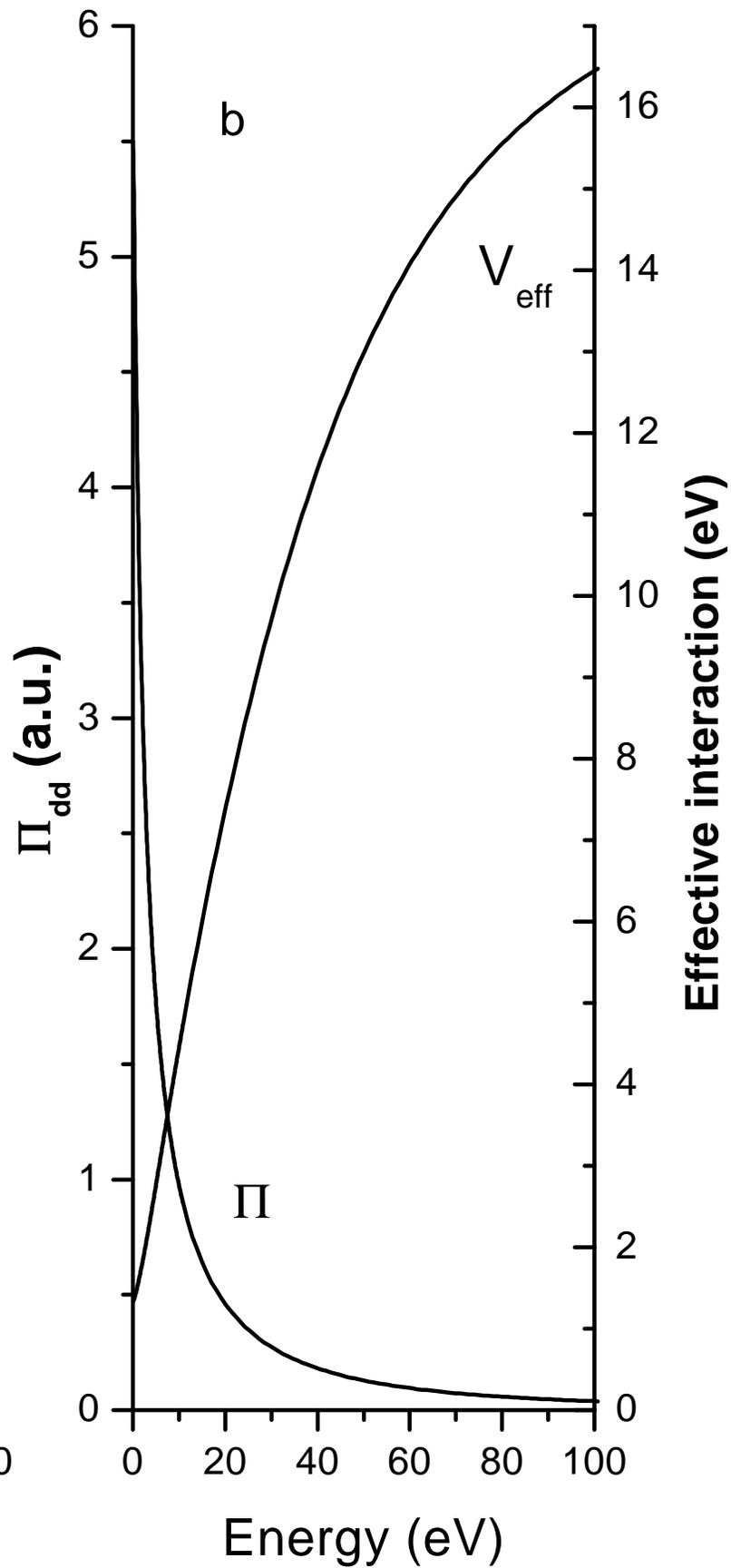
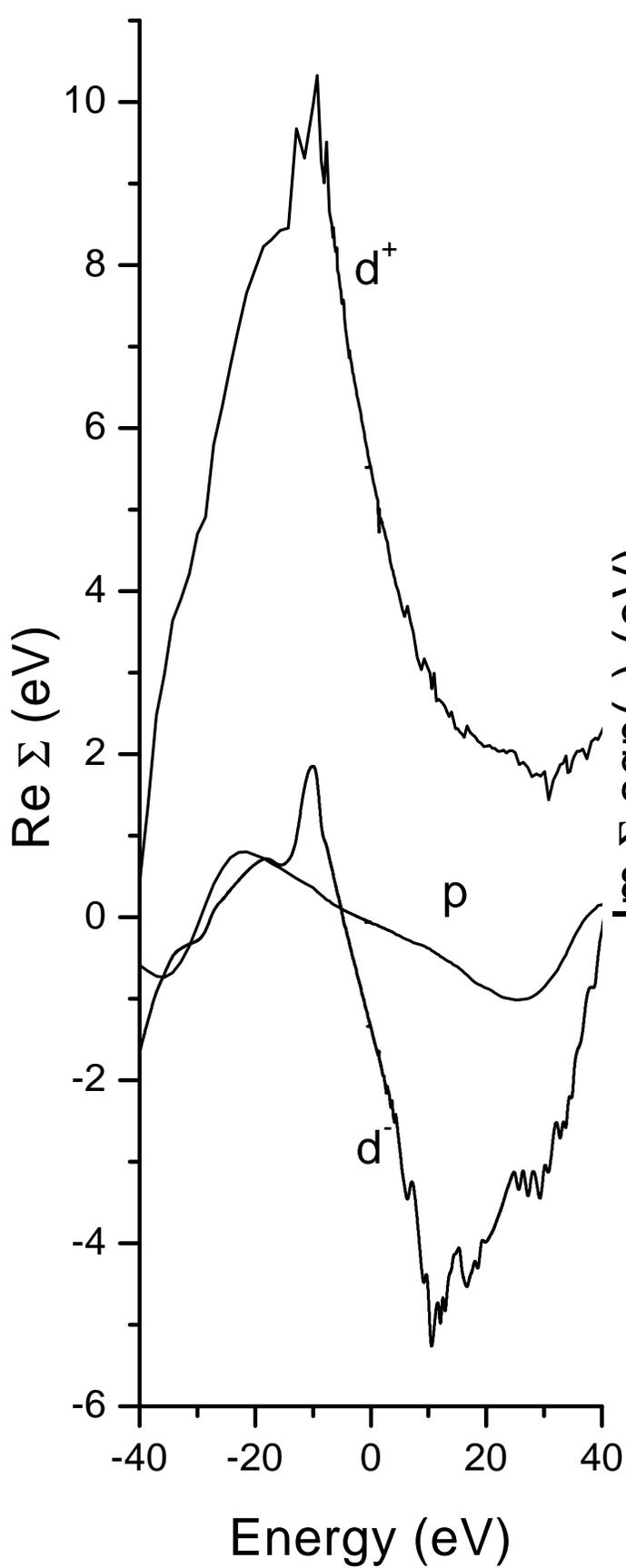
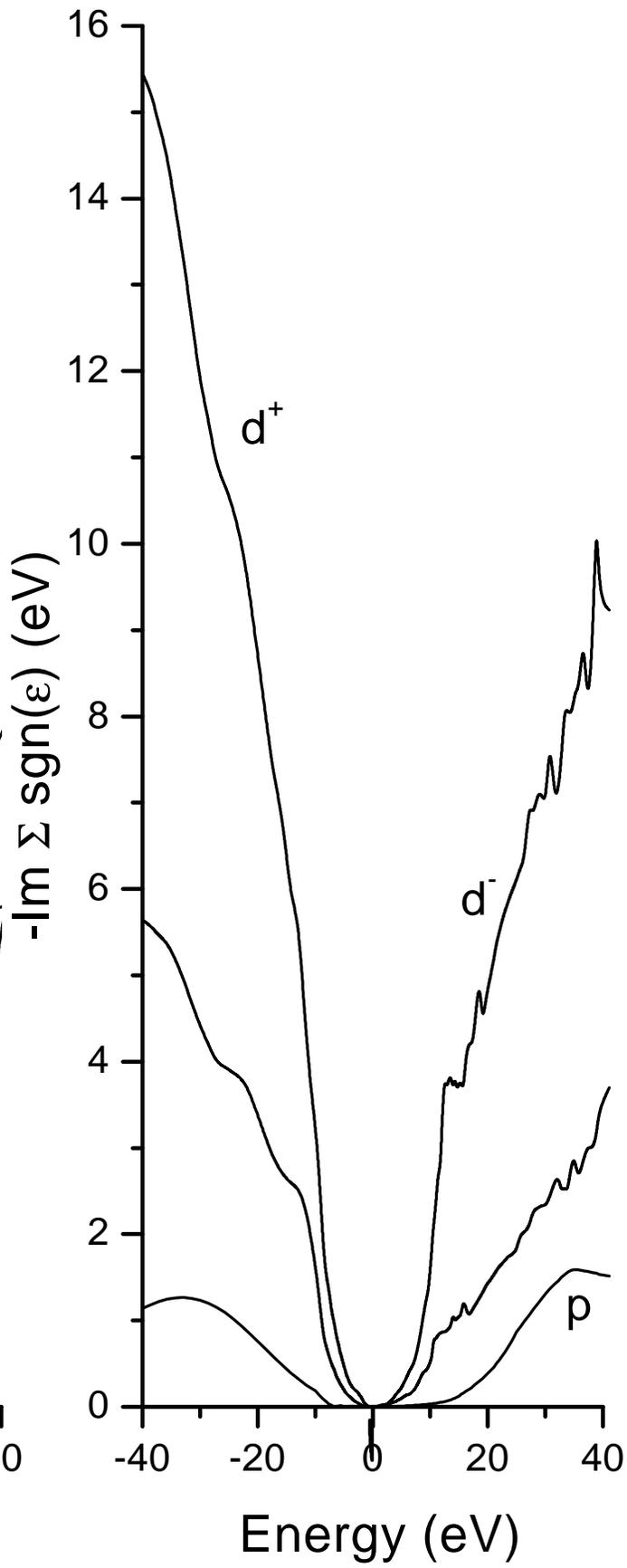
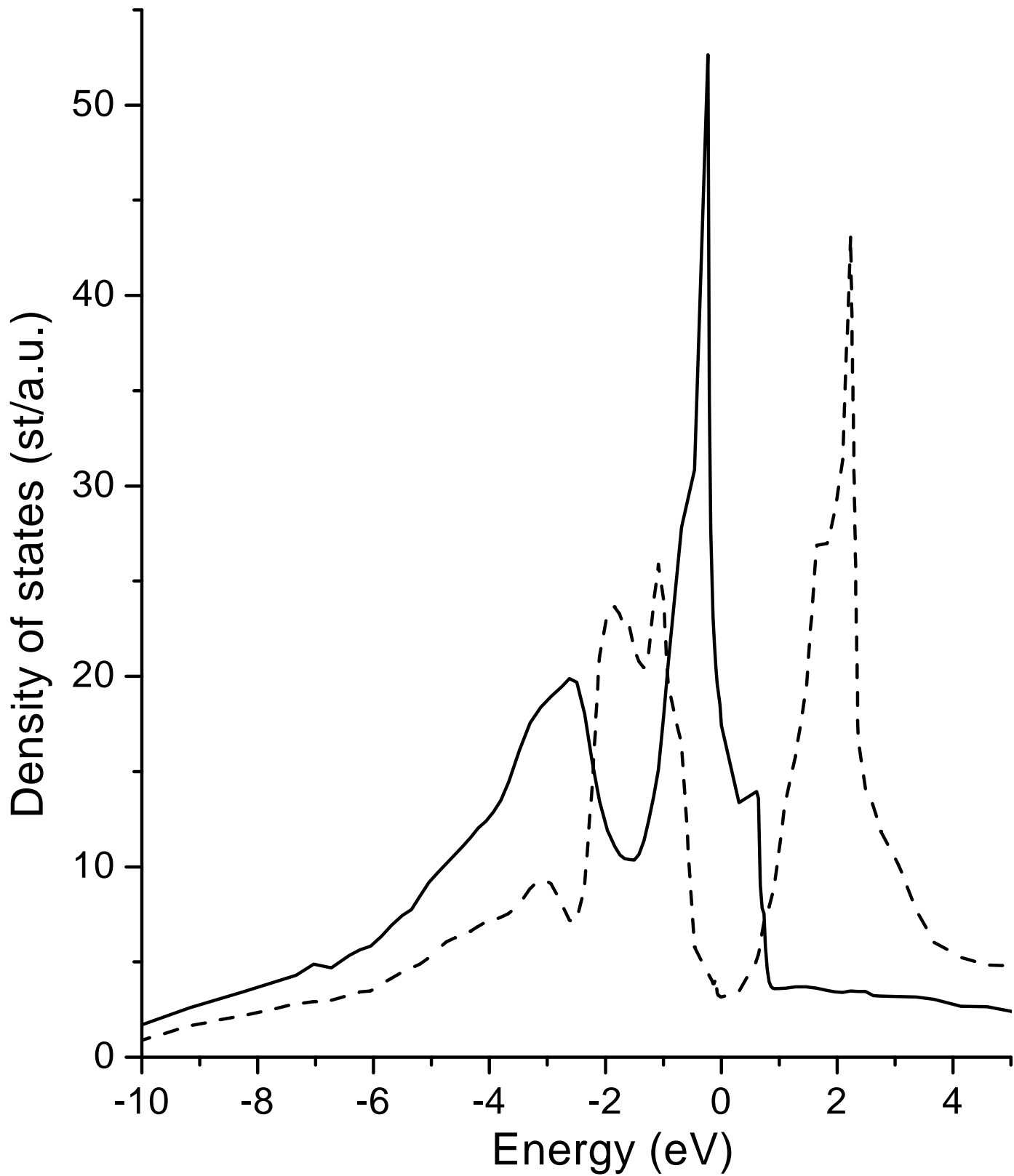
\end{document}